# Thermal properties of $3BaO-3TiO_2-B_2O_3$ glasses

Rahul Vaish and K.B.R. Varma*

Materials Research Centre,

Indian Institute of Science,

Bangalore – 560 012,

India.

*Corresponding Author; E-Mail : kbrvarma@mrc.iisc.ernet.in;

FAX: 91-80-23600683; Tel. No: 91-80-22932914

**Abstract:**

Transparent glasses in the system $3BaO$-$3TiO_2$-$B_2O_3$ (BTBO) were fabricated via the conventional melt-quenching technique. The as-quenched samples were confirmed to be glassy by differential thermal analysis (DTA) technique**.** The thermal parameters were evaluated using non-isothermal DTA experiments. Kauzmann temperature was found to be 759 K based on heating rate dependent glass transition and crystallization temperatures. Theoretical relation for temperature dependent viscosity was proposed for these glasses and glass-ceramics.

## 1. Introduction:

Glasses embedded with polar crystallites have been of increasing importance because of their multifarious properties that include electro-optic, ferroelectric, pyroelectric, piezoelectric etc. [1-4]. Indeed the physical properties of these glass-ceramics are dependent upon the size and shape of the crystallites, their volume fraction and connectivity. These parameters could be optimized using a controlled heat-treatment process. We have been fabricating various ferroelectric glass-ceramics from the point of view of exploiting them for various aforementioned device applications.

Borate-based single crystalline materials which include $Li_2B_4O_7$, $LiB_3O_5$, $BaB_2O_4$, $BiB_3O_6$, etc. were reported to exhibit interesting non-linear optic and related properties [5-8]. These crystals have wide transmission window, high optical damage threshold, good chemical and mechanical stability. We thought it was worth investigating into the physical properties of the glass-ceramics of these fascinating materials as this route facilitates to accomplish strict control over their microstructural characteristics and hence the physical properties.

$Ba_3Ti_3O_6(BO_3)_2$ crystals were reported to possess excellent non-linear optical properties [9, 10]. The optical second harmonic generation (SHG) efficiency of these crystals was about 95% of that of $LiNbO_3$ single crystals [10]. We have been making systematic attempts to fabricate glasses and glass-ceramics in the $BaO-TiO_2-B_2O_3$ system and visualize their physical properties. The composition for the present study is chosen in such a way that one could obtain $Ba_3Ti_3O_6(BO_3)_2$ crystalline phase on crystallization of $3BaO-3TiO_2-B_2O_3$ glasses. To begin with a detailed analyses of the glass-transition, crystallization kinetics and viscosity behavior of the glasses are in order. This would facilitate the optimization of certain parameters for obtaining transparent glass nano/microcrystals composites. The details pertaining to these studies are reported in the following sections.

**2. Experimental:**

Transparent glasses in the composition $3BaO-3TiO_2-B_2O_3$ (in molar ratio) were fabricated via the conventional melt-quenching technique. For this, $BaCO_3$, $TiO_2$ and $H_3BO_3$ were thoroughly mixed and melted in a platinum crucible at 1473K for 1h. The batch weight was 10 gm. Melts were quenched by pouring on a steel plate that was preheated at 423K and pressed with another plate to obtain 1-1.5 mm thick glass plates. All these samples were annealed at 773 K (5 h) which is well below the glass transition temperature. X-ray powder diffraction (Philips PW1050/37, Cu *K*a radiation) study was performed on the as-quenched powdered samples at room temperature to confirm their amorphous nature. The DTA (TA Instruments SDTQ 600) runs were carried out in the 573K –1173K temperature range. For non-isothermal experiments, the glass samples were heated from 573K to 1173K at the rate of 10, 20, 30, 35, 40 and 50K/min. Baseline (empty) and standard (indium) runs were also made in the same temperature range and at the same heating rates. The bulk glass samples of 1 mm thick, weighing 35 mg were taken in a platinum crucible and alumina powder was used as the reference for each experiment. All the experiments were conducted in dry nitrogen ambience.

## 3. Results and discussion

The DTA trace that was obtained for the as-quenched glass plates at a heating rate of 10 K/min is shown in Fig. 1. As expected, an endotherm followed by an exotherm were encountered for the as-quenched samples under study. The endotherm at 827 K corresponds to the glass-transition temperature ($T_g$). A subsequent exotherm that is incident around 920 K is attributed to the crystallization ($T_{cr}$) of the BTBO glasses. The X-ray powder diffraction (XRD) pattern of the as-quenched glasses confirms their amorphous nature (Fig. 2 (a)). The XRD pattern recorded for the sample heat-treated at the crystallization temperature (920 K) is shown in Fig. 2 (b). The Bragg peaks that are observed in this pattern are in good agreement with that reported in the literature for the poly-crystalline sample of $Ba_3Ti_3O_6(BO_3)_2$ [9] obtained by the solid-state reaction route.

The DTA traces that are obtained for the as-quenched glasses at different heating rates (10, 20, 30, 35, 40 and 50K/min) are shown in Fig. 3. The glass transition ($T_g$) and the crystallization ($T_{cr}$) temperatures shift towards higher temperatures with increasing heating rate (as illustrated in Fig. 3), indicating the kinetic nature of the glass transition and the crystallization. The heating rate dependent glass transition temperature is rationalized using fragmentation model [11]. According to this model, viscous behavior is caused by the fragmentation of random networks due to bond breaking. The density of the broken bonds could be obtained from the rate equation which is dependent on the heating rate [11]. The fragmentation would occur when the density of the broken bonds reaches a certain level. The system comprising fragments must exhibit a lower viscosity than that of the continuous random networks, and the viscosity of the fragmentized system decreases as the fragmentation proceeds. Therefore, the glass transition is dependent on the heating rate, as the fragmentation processes are heating rate dependent. The investigation into the heating rate dependent glass-transition and crystallization could be helpful for the understanding of thermal behavior of BTBO glasses.

The effect of heating rate ($a$) on $T_g$ and $T_{cr}$ has been rationalized using the Lasocka's relation [12]. The relation between the $T_g$, $T_{cr}$ and $a$ could be represented empirically (Lasocka's relation) as follows:

$$T_g = A_g + B_g \log\alpha \qquad (1)$$

and

$$T_{cr} = A_{cr} + B_{cr} \log\alpha \qquad (2)$$

where $A_g$ and $A_{cr}$ are the glass transition and crystallization temperatures for the heating rate of 1 K/min. $B_g$ and $B_{cr}$ are the constants. The plots of $T_g$ and $T_{cr}$ versus log $a$ for BTBO glass-plates along with the theoretical fits (solid lines) are shown in Fig. 4. The above relations (Eqs. 1 and 2) can be written as;

$$T_g \text{ (K)} = (811 \pm 2) + (16 \pm 1) \log a \text{ (K/min)} \qquad (3)$$

and

$$T_{cr} \text{ (K)} = (882 \pm 1) + (38 \pm 1) \log a \text{ (K/min)} \qquad (4)$$

The crystallization has a stronger dependence on heating rate than that of the glass transition ($B_{cr} > B_g$). The relative location of $T_{cr}$ with reference to $T_g$ in the DSC/DTA trace is a measure of the thermal stability of glasses. The thermal stability ($T_{cr}$-$T_g$) of the glasses is a crucial parameter to be noted from their technological applications point of view. Glasses should be sufficiently stable against crystallization. However, one must be able to form nuclei and subsequently grow crystals within the glass matrix on heat treatment inorder to obtain glass-ceramics. The higher values of the $T_{cr}$-$T_g$ delay the nucleation process. The parameter ($T_{cr}$-$T_g$) is heating rate dependent and on combining Eqs. 1 and 2, one arrives at:

$$T_{cr}-T_g = (A_{cr}-A_g) + (B_{cr}- B_g) \log a \quad (5)$$

The above relation for BTBO glass-plates can be written as

$$T_{cr}-T_g \text{ (K)} = (71 \pm 2) + (22 \pm 1) \log a \text{ (K/min)} \quad (6)$$

Using the above equation, the values of $T_{cr}$ –$T_g$ at various heating rates could be obtained. It is clear that $T_{cr}$-$T_g$ decreases with decreasing heating rate (Eq. 6) and can be equal ($T_{cr} = T_g$) at some lower value of heating rate $a_k$, which can be expressed as;

$$\log a_k = - [(A_{cr}-A_g)/(B_{cr}-B_g)] \quad (7)$$

The $a_k$ value that is obtained for BTBO glass-plates is 0.0006 K/min. It is to be noted that at this heating rate ($a_k$), glass is expected to undergo the glass-transition and the crystallization processes simultaneously. This is known as the Kauzmann temperature ($T_k$) [13-15]. The calculated value for $T_k$ for the present glass system is 759 K. It is experimentally difficult to observe this ideal glass-transition temperature ($T_k$) as the heating rate required is extremely low (0.0006 K/min). The estimated time involved is about 531 days to achieve the above temperature ($T_k$) at that heating rate.

It is of interest to investigate the activation energy of the glass-transition for the BTBO glass samples. The activation energy ($E_g$) associated with the glass-transition is involved in the molecular motion and rearrangement of atoms. The activation energy for the glass-transition for the present glasses could be evaluated using Kissinger's method [16] and is given by the following relation:

$$\ln (a/T_g^2) = -E_g/RT_g + \text{constant} \quad (8)$$

where R is the universal gas constant. The above equation (Eq. 8) was formerly derived for the process of crystallization. However, it is known in the literature [17] that the Kissinger equation could also be used to evaluate the activation energy associated with the glass transition. A plot of $\ln(a/T_g^2)$ versus $(1000/T_g)$ yields a linear relation which is depicted in Fig. 5. The experimental points of the present work along with a linear fit (solid line) to the above relation suggest its validity. From the slope of the straight line, the value of $E_g$ is found to be 740 ± 10 kJ/mol. The variation in $\ln T_g^2$ with $a$ is insignificant as compared to the change in $\ln a$ [18]. Therefore, the eq. 8 could be written as;

$$\ln a = - E_g/RT_g + \text{constant} \qquad (9)$$

The value of $E_g$ is obtained from the slope of a straight line (linear fit) of ln a vs $1/T_g$ plot. The plots of $\ln a$ vs $1000/T_g$ along with linear fits are shown in Fig. 5. The average value of $E_g$ obtained from the slope of the linear fit is 735 ± 8 kJ/mol which is in close agreement with that obtained by Kissinger's method.

Another important parameter of the glasses is their Fragility index ($F$) which is a measure of the rate at which the relaxation time decreases with increasing temperature around $T_g$. The glass forming liquids that exhibit an approximately Arrhenius temperature dependence of the viscosity are defined as strong glass formers and those which exhibit a non-Arrhenius behavior (Vogel Fulcher Tammann (VFT) equation [19,20]) are known as fragile glass formers. The value of the fragility index ($F$) could be estimated using the following relation [21]:

$$F = \frac{E_g}{RT_g \ln 10} \qquad (10)$$

It was known in the literature that kinetically strong-glass forming liquids have low values of $F$ ($F \approx 16$), while fragile glass forming liquids have higher values of $F$ ($F \approx 200$). The value of $F$ obtained for the glasses under investigation at the heating rate of 10K/min is 47 ± 2. This value which is well within the above mentioned range indicates that the present system can be considered to belong to a strong glass forming family.

It is essential to have apropri knowledge about the crystallization behavior of the BTBO glasses for fabricating the glass-ceramics with the desired characteristics. Crystallization behavior of the BTBO glass samples was studied by non-isothermal methods. For evaluating the activation energy ($E_{cr}$) for crystallization using the variation in $T_p$ with $a$, Kissinger [16] developed a method for a non-isothermal analysis of the crystallization which is as follows:

$$\ln\left(\frac{\alpha}{T_p{}^2}\right) = -\ln\left(\frac{E_c}{R}\right) + \ln K_o - \frac{E_{cr}}{RT_p} \qquad (11)$$

The plots of $\ln(\alpha / T_P^2)$ versus $1000/T_p$ for these glasses are shown in Fig. 6. The value of the activation energy and the frequency constant ($K_o$) are obtained from the slope ($E_{cr}/R$) and intercept ($\ln K_0 - \ln\left(E_{cr}/R\right)$) of the linear fit to the experimental points. The value of $E_{cr}$ is 350 ± 10 kJ/mol. for the glasses understudy. The value for the frequency constant is about 1.6 X $10^{19}$ $sec^{-1}$ which indicates the number of attempts per second made by the nuclei to overcome the energy barrier.

In order to have further insight into the thermal behavior of BTBO glasses, its temperature-dependent behavior of the viscosity (?) is studied using VFT equation. It is assumed that the temperature dependent viscosity for the BTBO glasses could be expressed by VFT relation as follows

$$\eta = \eta_o \exp\left[ B / (T - T_o) \right] \tag{12}$$

where ?$_o$ and $B$ are constants. And $T_o$ is the VFT temperature corresponding to the kinetic instability point [22]. Chen [23] showed that the activation energies associated with the glass transition and the crystallization could be associated with those for viscous flow as a function of temperature;

$$E(T) = R\frac{d(\ln\eta)}{d(1/T)} = \frac{RBT^2}{(T - T_o)^2} \tag{13}$$

The above equation (Eq. 13) for the glass-transition and the crystallization could be written as;

$$E_g = \frac{RBT_g^2}{(T_g - T_o)^2} \tag{14}$$

and for the crystallization,

$$E_{cr} = \frac{RBT_{cr}^2}{(T_{cr} - T_o)^2} \tag{15}$$

After solving Eqs. 14 and 15 for $B$ and $T_o$, one would arrive at;

$$B = \left(\frac{E_g}{C}\right) * \left(\frac{T_g - T_{cr}}{T_g}\right)^2 \tag{16}$$

and

$$T_o = \left[\frac{T_g\sqrt{C} - T_{cr}}{\sqrt{C} - 1}\right] \tag{17}$$

where $C = \left( \frac{E_g}{E_{cr}} \right) \left( \frac{T_{cr}}{T_g} \right)^2$ (18)

The obtained values for $E_g$ and $E_{cr}$ were used to deduce the constants ($B$ and $T_o$) in the above relations (Eqs. 16 and 17). For this purpose, the values of $T_g$ and $T_{cr}$ at 10 K/min were considered. The values obtained for $B$ and $T_o$ are around 5180 K and 676 K, respectively. It is to be noted that the value of $T_o$ is less than the that of Kauzmann temperature, ($T_k$.) It is reported in the literature that the relation $T_k = T_o$ is valid only for fragile glasses whereas for strong glasses $T_k > T_o$ [22].

To establish the theoretical relation for temperature dependence of viscosity, it is considered that a liquid on cooling becomes a glass when the viscosity equals $10^{12}$ poise ($?(T_g)$ = $10^{12}$ poise). The numerical value for $?_o$ could be deduced from Eq. 12 using the values for $B$, $T_o$ and $?(T_g)$ ($10^{12}$ poise). The expression for the viscosity (Eq.12) for the BTBO glasses may be written as;

$$\eta = 1.23 \times 10^{-3} \exp\left[ 5180 / (T - 676) \right] \quad (19)$$

The viscosity at various temperatures for the BTBO glass in the supercooled liquid region can be estimated using the Eq. 19. The value of viscosity is expected to increase with the increase in the volume fraction of crystallization. The viscosity expression for the glass-ceramic of various fractions of crystallization is [24];

$$\frac{\eta_{eff}}{\eta} = (1 + mx)^n \quad (20)$$

where $\eta_{eff}$ is the viscosity for the crystallized glass with volume fraction, $x$ and ? is the viscosity of the as-quenched glasses. The constants $m$ and $n$ are dependent on the shape and orientation of the crystallites in the glass matrix. Under the assumption that spherical crystals are randomly oriented in the glasses, Eq. 20 could be written as

$$\frac{\eta_{eff}}{\eta} = (1 - x)^{-3} \qquad (21)$$

The above equation is valid only for the value of $x$ lying in the range 0-0.6. One can obtain the viscosities at various temperatures for the BTBO glass-ceramics using the Eqs. 19 and 21. The plots of viscosity versus temperature for the glasses and glass-ceramics ($x$=0.3 & 0.6) are depicted in Fig. 7. It is clear from the figure that the viscosity increases with increase in the volume fraction of crystallization and decreases with increase in temperature.

## 4. Conclusions:

The glass-transition and crystallization kinetics of $3BaO\text{-}3TiO_2\text{-}B_2O_3$ glasses have been studied. The activation energies associated with the glass-transition and crystallization were used to determine the temperature dependent viscosity relation (VFT equation) and is expressed as $\eta = 1.23\times10^{-3}\exp\left[5180/(T - 676)\right]$. The value for Kauzmann temperature is 759 K. The value obtained for the fragility index indicates that the present composition belongs to a strong glass forming family of oxides. The thermal parameters that are obtained in the present studies are crucial for controlling the crystallization in bulk samples.

**Figure captions:**

Fig. 1: DTA trace for as-quenched $3BaO-3TiO_2-B_2O_3$ glass-plates

Fig. 2: X-ray diffraction patterns for the (a) as-quenched and (b) crystallized powdered samples [$Ba_3Ti_3O_6(BO_3)_2$]

Fig. 3: DTA traces for as-quenched $3BaO-3TiO_2-B_2O_3$ glass-plates at various heating rates

Fig. 4: $T_g$ and $T_{cr}$ versus log $a$ for $3BaO-3TiO_2-B_2O_3$ glass-plates

Fig. 5: ln $a$ and ln $(a/T_g^2)$ versus $1000/T_g$

Fig. 6: ln $(a/T_P^2)$ versus $1000/T_P$ plot

Fig. 7: ln ? & ln $?_{eff}$ versus $T$ plots for glasses and glass-ceramics of two different fractions of crystallization

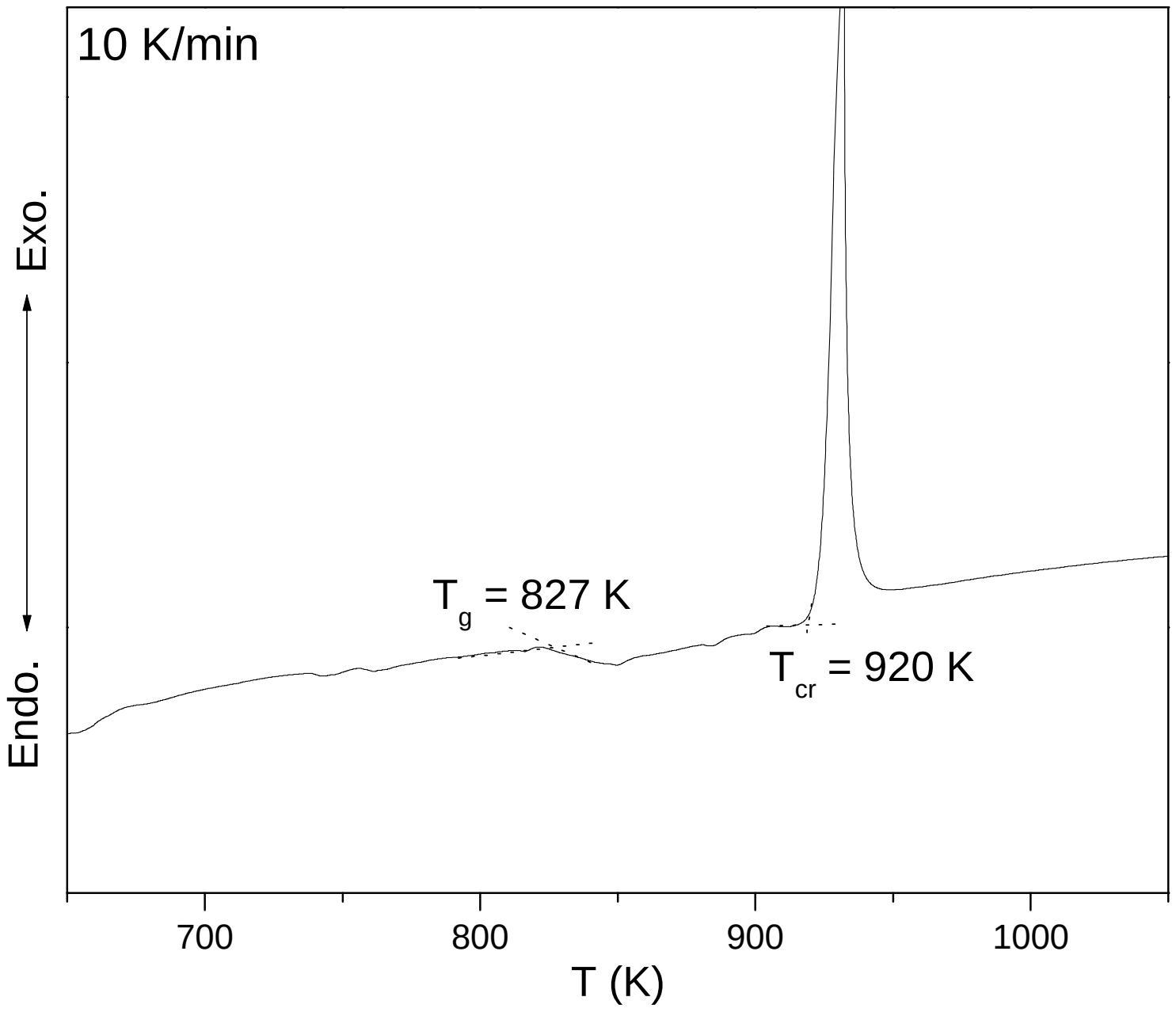

Fig.1

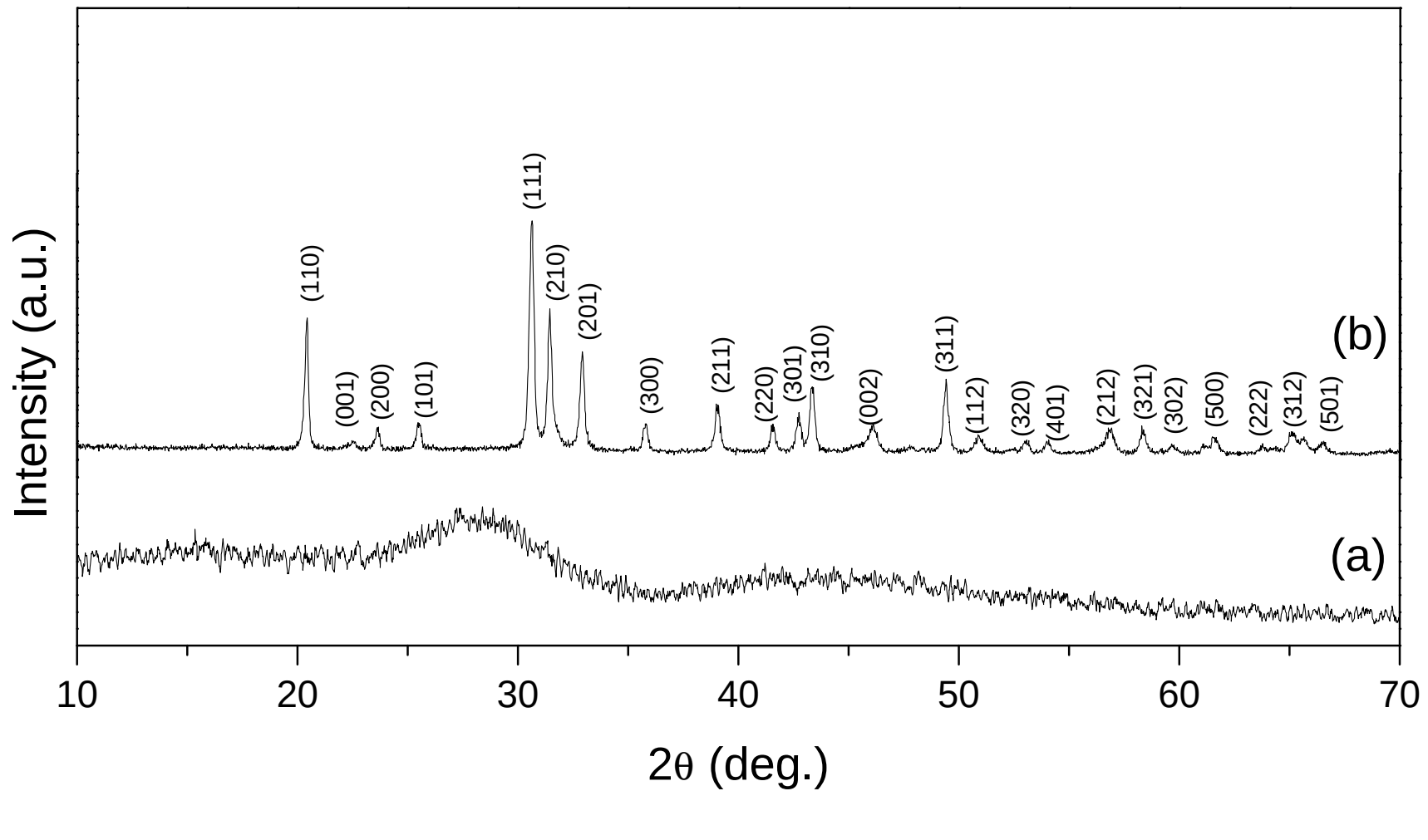

Fig. 2

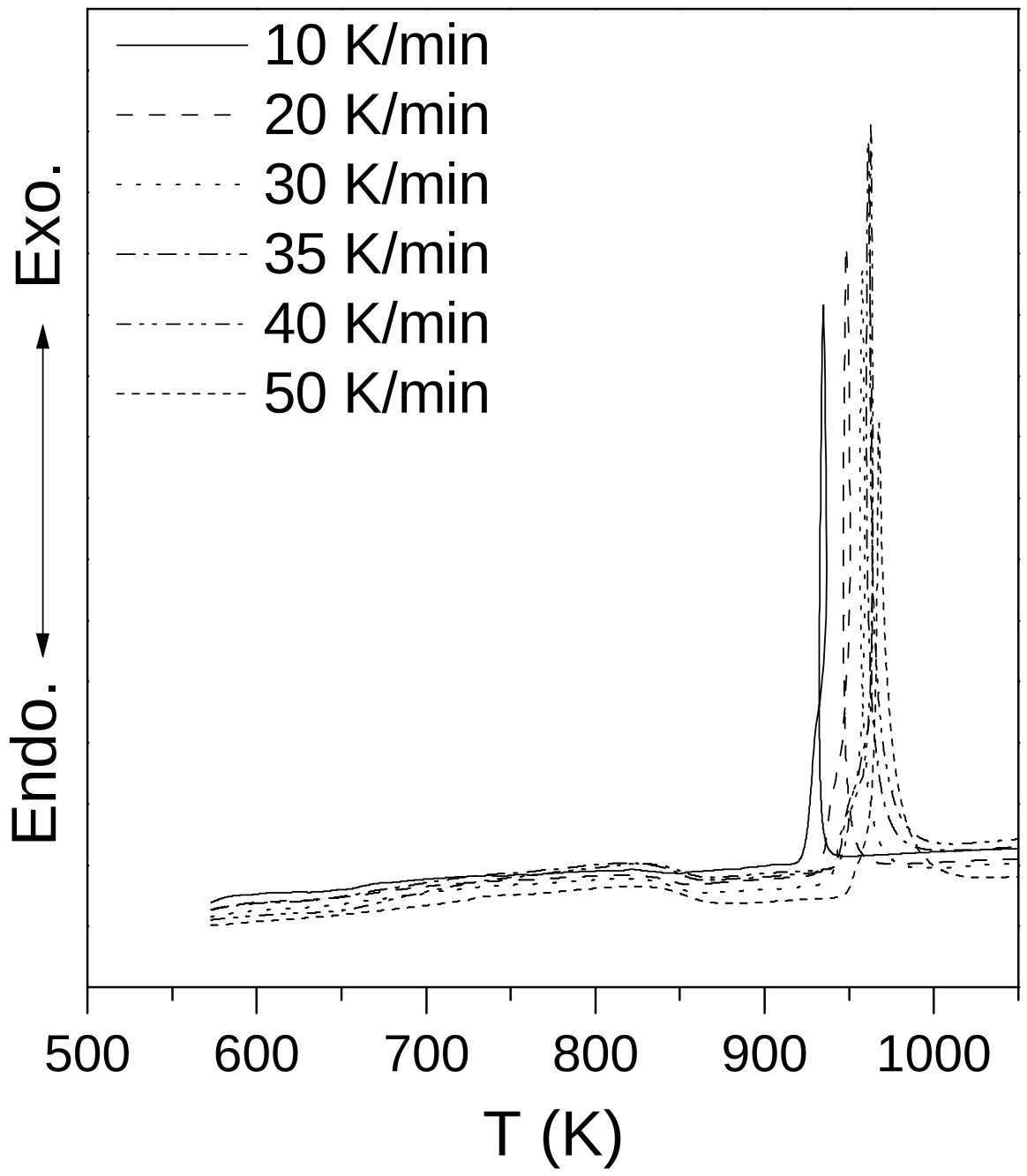

Fig. 3

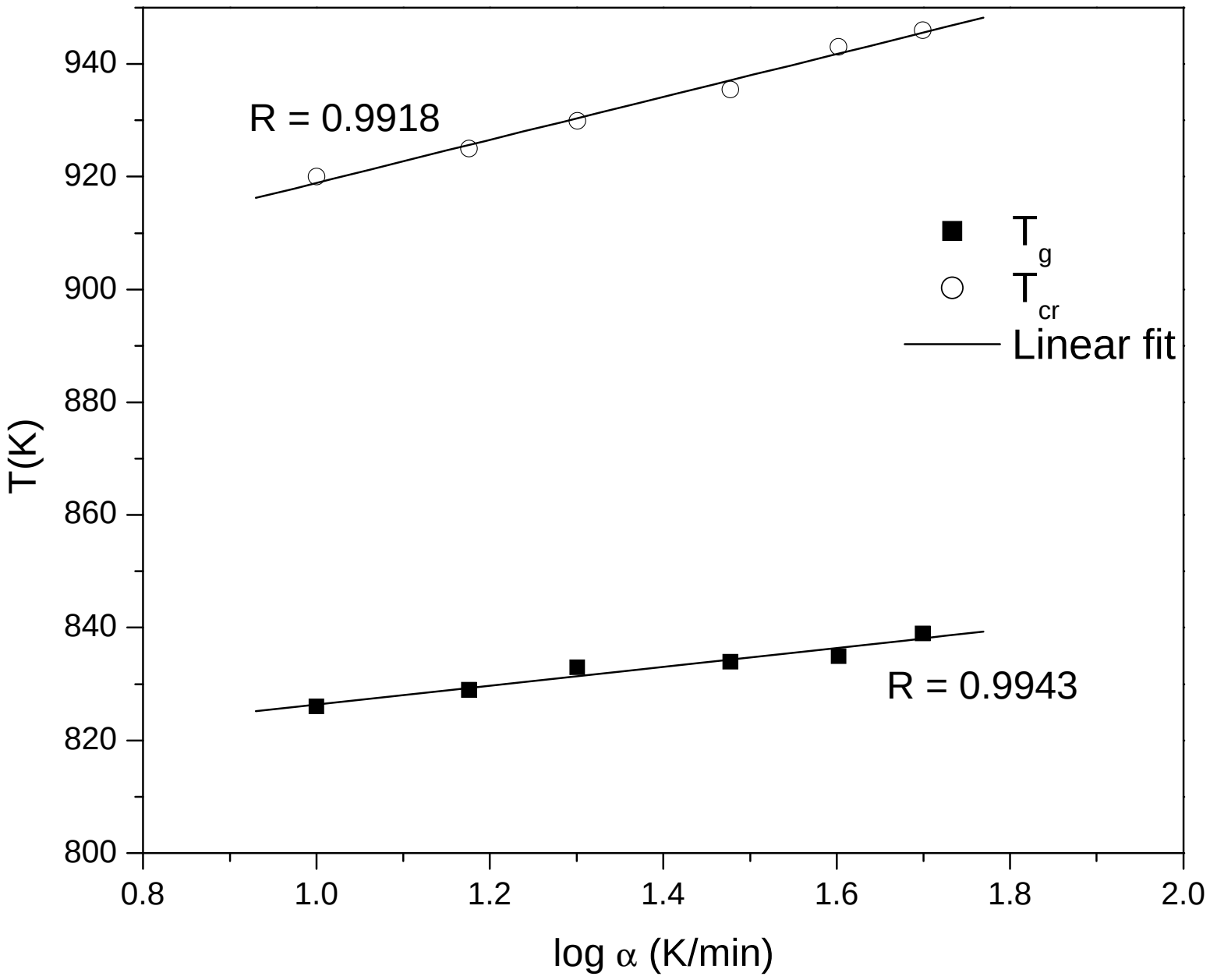

Fig. 4

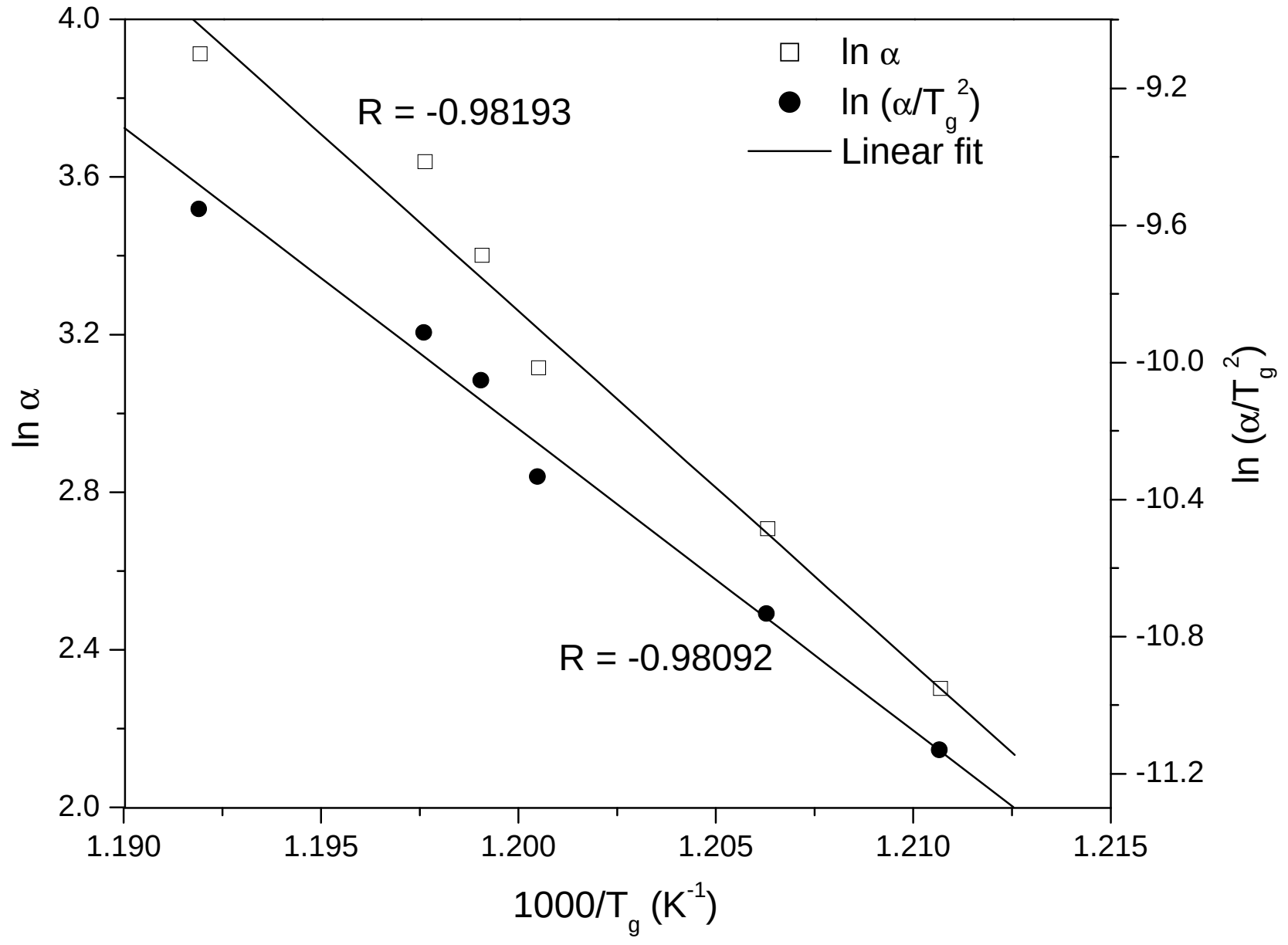

Fig. 5

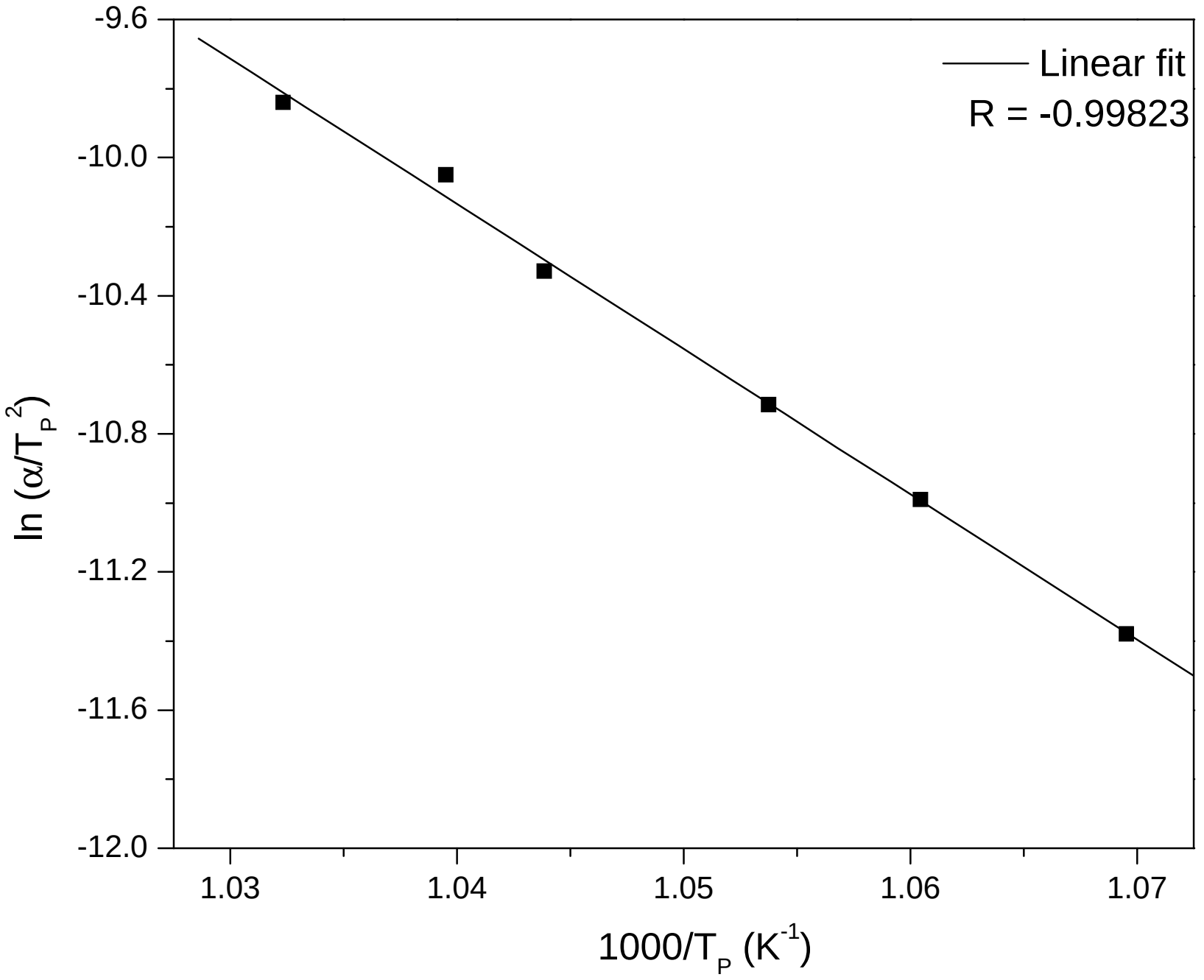

Linear fit
R = -0.99823
ln (α/T_P^2)
1000/T_P (K^-1)

Fig. 6

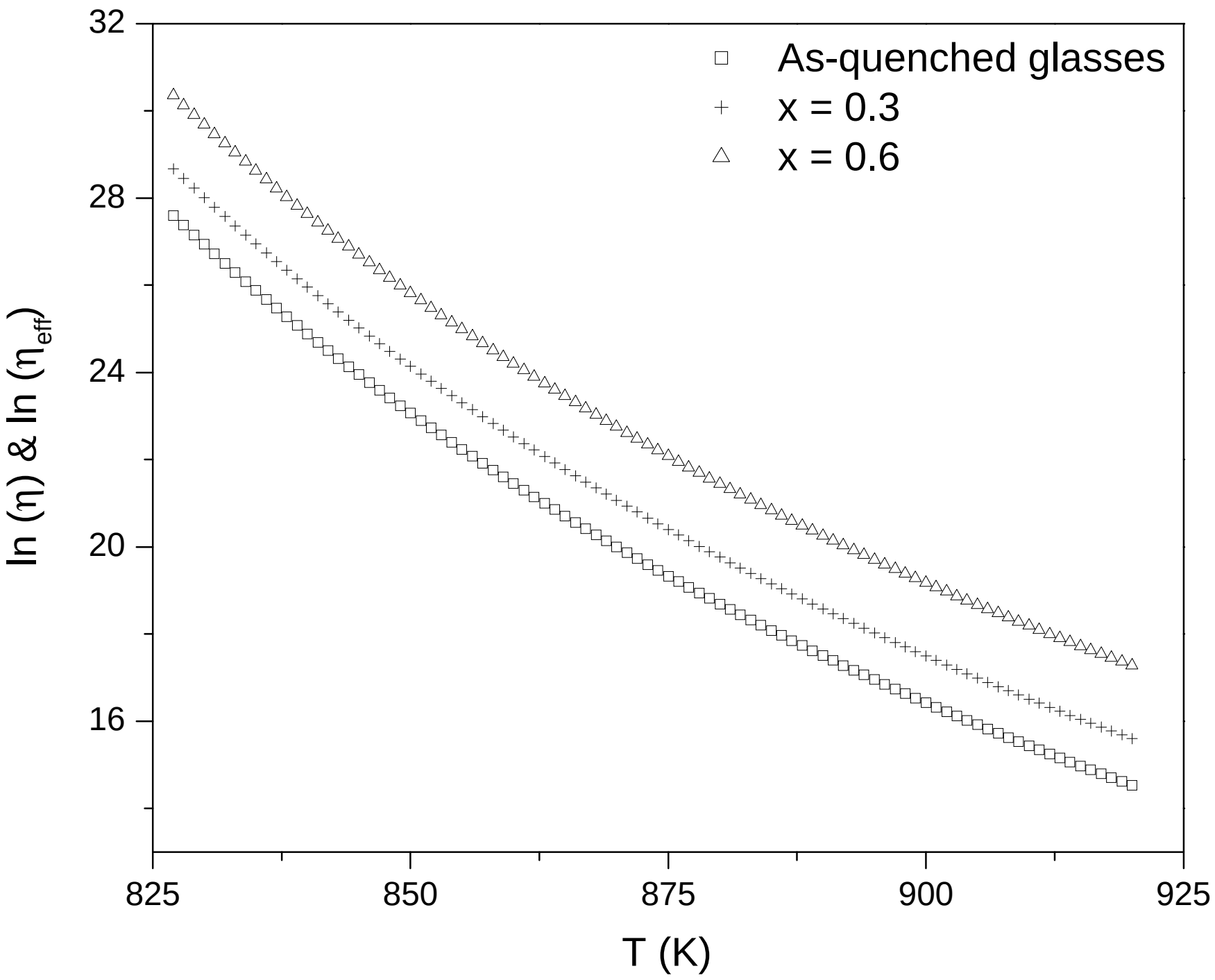

Fig. 7